\documentclass[a4paper,11pt]{article}
\usepackage{pos}
\usepackage{graphicx} 
\usepackage{amsmath}
\newcommand{\Dipper}{\textsc{McDipper}}

\newcommand{\rmd}{\mathrm{d}}

\newcommand{\xT}{\mathbf{x}}

\newcommand{\pT}{\mathbf{p}}

\definecolor{oscar}{RGB}{22, 156, 172}

\title{Charge and energy deposition in the \Dipper{} framework}

\author*[a]{Oscar Garcia-Montero}
\author[a]{Sören Schlichting}
\author[b,c,d,e]{Hannah Elfner}

\affiliation[a]{Fakult\"at f\"ur Physik, Universit\"at Bielefeld,\\
	D-33615 Bielefeld, Germany}

\affiliation[b]{GSI Helmholtzzentrum f\"ur Schwerionenforschung,\\ Planckstr. 1, 64291 Darmstadt, Germany}

\affiliation[c]{Institute for Theoretical Physics, Goethe University,\\
	Max-von-Laue-Strasse 1, 60438 Frankfurt am Main, Germany}

\affiliation[d]{Frankfurt Institute for Advanced Studies,\\ Ruth-Moufang-Strasse 1, 60438
	Frankfurt am Main, Germany}

\affiliation[e]{Helmholtz Research Academy Hesse for FAIR (HFHF), GSI Helmholtz Center,
	Campus Frankfurt\\
	Max-von-Laue-Straße 12, 60438 Frankfurt am Main, Germany}

\emailAdd{garcia@physik.uni-bielefeld.de}

\notes{\note{\textbf{Acknowledgment}:  The authors acknowledge support by the Deutsche Forschungsgemeinschaft (DFG, German Research Foundation) through the CRC-TR 211 ‘Strong-interaction matter under extreme conditions’-project number 315477589 – TRR 211. OGM and SS acknowledge also support by the German Bundesministerium für Bildung und Forschung (BMBF) through Grant No. 05P21PBCAA.}}

\abstract{In this short note we present aspects of the energy and charge deposition within the \Dipper{}, a novel 3D resolved model for the initial state of ultrarelativistic Heavy-Ion collisions based on the $k_\perp$-factorized Color Glass Condensate hybrid approach. This framework is a initial-state Monte Carlo event generator which deposits the relevant conserved charges (energy, charge and baryon densities) both in the midrapidity and forward/backward regions of the collision. The event-by-event generator computes the gluon and (anti-) quark phase-space densities using the IP-Sat model, from where the conserved charges can be extracted directly. In this work we present the centrality and collision energy dependence for the deposited conserved quantities at midrapidity and the full event, the so-called $4\pi$ solid angle range. }

\FullConference{%
	HardProbes 2023,\\
	26-31 March 2023\\
	Aschaffenburg, Germany
}

\begin{document}
\maketitle

\section{Introduction}

It is widely accepted that during a Heavy Ion Collision (HICs) the formation of a complex, hot and dense system comprised of quarks and gluons is formed, the so-called Quark-Gluon Plasma. In recent years, an interest has risen over observations of a rich longitudinal structure of the long-range correlations of the system~\cite{CMS:2015xmx,Nie:2019bgd,Petersen:2011fp,Pang:2018zzo}. Such structure is thought to be created through diverse mechanisms during the initial instants, and translated to momentum space through interactions. 
Therefore,  new experimental and theoretical insights towards the forward/backward rapidity windows of heavy ion collisions call for a fast and phenomenologically available initial conditions model with non-trivial rapidity dependence and strongly inspired by first principles computations. Motivated by these ideas, as well as by the need for a  framework for comprehensive comparison of saturation physics and models in HICs  and the upcoming Electron-Ion Collider, we recently presented the \Dipper{} framework~\cite{Garcia-Montero:2023gex}. This saturation based frameworks provides a flexible and systematically improvable 3D initial conditions for HICs, firmly rooted in high energy Quantum Chromodynamics.

Expanding on our recent work,  we present a study on the initial charge and energy deposition,  valuable for many aspects in high energy nuclear physics, ranging from the initial dynamics of the pre-equilibrium stage, the underlying understanding of net-particle yields and correlations and the production of electromagnetic,  among others. We will focus on the deposition at midrapidity of energy, $\rmd E/\rmd \eta_s$,  electric charge, $\rmd Q/\rmd \eta_s$, and baryon charge , $\rmd B/\rmd \eta_s$. We will compare this to the event total, i.e. the rapidity integrated-charges. In what follows, we will call this rapidity integrated charges the full solid angle or $4\pi$ charges, e.g. $Q_{4\pi}\equiv \int \rmd \eta_s \rmd Q/\rmd \eta_s$. 
 
\section{The \Dipper{} framework}

The \Dipper{} framework computes the initial energy and charge deposition in high-energy HICs within the dilute-dense approximation of the $k_\perp$ factorized  Color Glass Condensate Effective Field Theory~\cite{Gribov:1983ivg}. Using the  single particle production formulas, one can compute inclusive gluon $(\rmd N_{g}/{\rmd^2\xT \rmd^2\pT \rmd y})$ and (net-) quark distributions $\rmd N_{\bar{q}-q}/{\rmd^2\xT \rmd^2\pT \rmd y}$ as a function of (momentum) rapidity $y$ transverse momentum $\pT$ and transverse position $\xT$. This is done by evaluating the leading order cross-sections for the transverse momentum dependent dipole gluon distributions from the IPSat model~\cite{Bartels:2002cj,Kowalski:2003hm}  and collinear parton distributions (PDFs) from the LHAPDF library~\cite{Buckley:2014ana}. For details on the computation of the gluon and quark single particle distributions please see the main reference, ref. \cite{Garcia-Montero:2023gex}. The relevant macroscopic conserved quantities can be calculated by taking moments of the single particle distributions and noting that at the LO in the high energy limit, the indentification of the momentum and spacetime rapidities can be taken, $y=\eta_s$. Energy deposition in this framework obeys the relation
\begin{equation}
	(e\tau)_0 =\int \rmd^2\pT~|\pT|~\left[K_g \frac{\rmd N_{g}}{\rmd^2\xT \rmd^2\pT \rmd y} + \sum_{f,\bar{f}} \frac{\rmd N_{q_f}}{\rmd^2\xT \rmd^2\pT \rmd y}\right]_{y=\eta_s}\;,
	\label{eq:EnergyWithKFactor}
\end{equation}
where the prefactor $K_g$ is a free paramter which accounts for the uncertainties coming from higher order perturbative corrections to the total cross-sections. Notice that since the IP-Sat model have been fitted to HERA data~\cite{Rezaeian:2012ji}, the $K_g$ prefactor in the gluon energy is the only free parameter in the framework and is fixed by fitting to the deposited energy of $pp$ and $pPb$ collisions at minimum bias for some collisional center-of-mass energy per nucleon,$ \sqrt{s_{\rm NN}}=5.02$. We choose to fix at  $ \sqrt{s_{\rm NN}}=5.02$ TeV, leading to $K_g=1.85$ for the IP-Sat model.
On the other hand, the quark number densities $n_f$ with $f=(u,d,s)$, can be computed using
\begin{equation}
	(n_f\tau)_0 =\int \rmd^2\pT~\left[\frac{\rmd N_{q_f}}{\rmd^2\xT \rmd^2\pT \rmd y} - \frac{\rmd N_{\bar{q}_f}}{\rmd^2\xT \rmd^2\pT \rmd y}\right]_{y=\eta_s} 
	\label{eq:Charges}
\end{equation}

\begin{figure}
 	\centering
 	\includegraphics[width=0.96\linewidth]{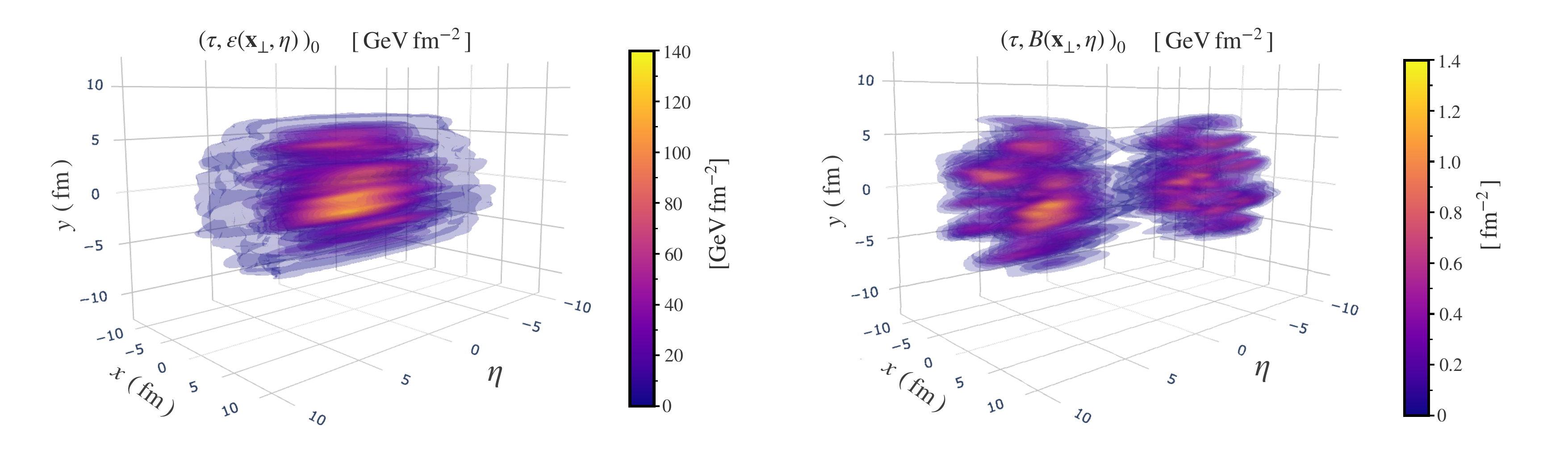}
 	\caption{Full spatial resolution for the deposited energy (\textit{left}) and baryon charge  (\textit{right}) of a single event of a Pb-Pb collision  at$ \sqrt{s_{\rm NN}}=5.02$~TeV and impact parameter $b=2$~fm. The computation is done for the IP-Sat model with the CT18NNLO parton distribution set~\cite{Bartels:2002cj,Kowalski:2003hm,Buckley:2014ana}.}
 	\label{fig:promo}
 \end{figure}

The electric charge density is then computed as $(\tau Q)_0 = \sum_f Q_f \, n_f$  where $Q_f= -1/3$ for $f=d,s$ quarks and $Q_f= 2/3$ for the $u$ quark. while the baryon charge is given by $(\tau B)_0 = Q_B\sum_f  \, n_f$, with $Q_B=1/3$ the baryon charge of a valence quark. The framework does not yet include spatial charge fluctuation coming from fluctations in the PDFs, and therefore the strangeness density vanishes trivially. With these conditions, a typical event in the \Dipper{} framework can be pictured in Fig.~\ref{fig:promo}, where energy (\text{left}) and baryon charge (\text{right}) are pictured for a single event at impact parameter $b=2\,$fm for a Pb-Pb collision at  $ \sqrt{s_{\rm NN}}=5.02$ TeV. As one can expect, the energy density is symmetric and higher at midrapidity. As we will see in the following section this is thanks to the gluon domination of the energy density at high collisional energies close to $\eta_s=0$. The interesting visualization is the right hand panel where we find the baryon density, which, as expected for a high energy event, is very small at midrapidity and peaks in the forward/backward regions. The charge distributions are in general asymmetric, thanks to their dependence on the nuclear density. As has been more extensively explained in ref.~\cite{Garcia-Montero:2023gex}, this dependence is not trivial, but one can schematically think of it roughly as $T_1 T_2^{a} + T_2 T_1^a$, where$T_{1/2}$ denotes the nuclear thickness, and $a\sim 1/2$.  Additionally, if one wants to find the charge and energy yields, e.g. the baryon charge yield, one needs to use $(\tau B)_0 = \rmd B /\rmd^2 \xT\rmd \eta_s $, which means that $B_{4\pi} = \int \rmd ^2 \xT\, \rmd \eta_s (\tau B)_0 $. The same is true for the energy and electric charge densities.

\section{Results}

\begin{figure}
	\centering
	\includegraphics[width=1\linewidth]{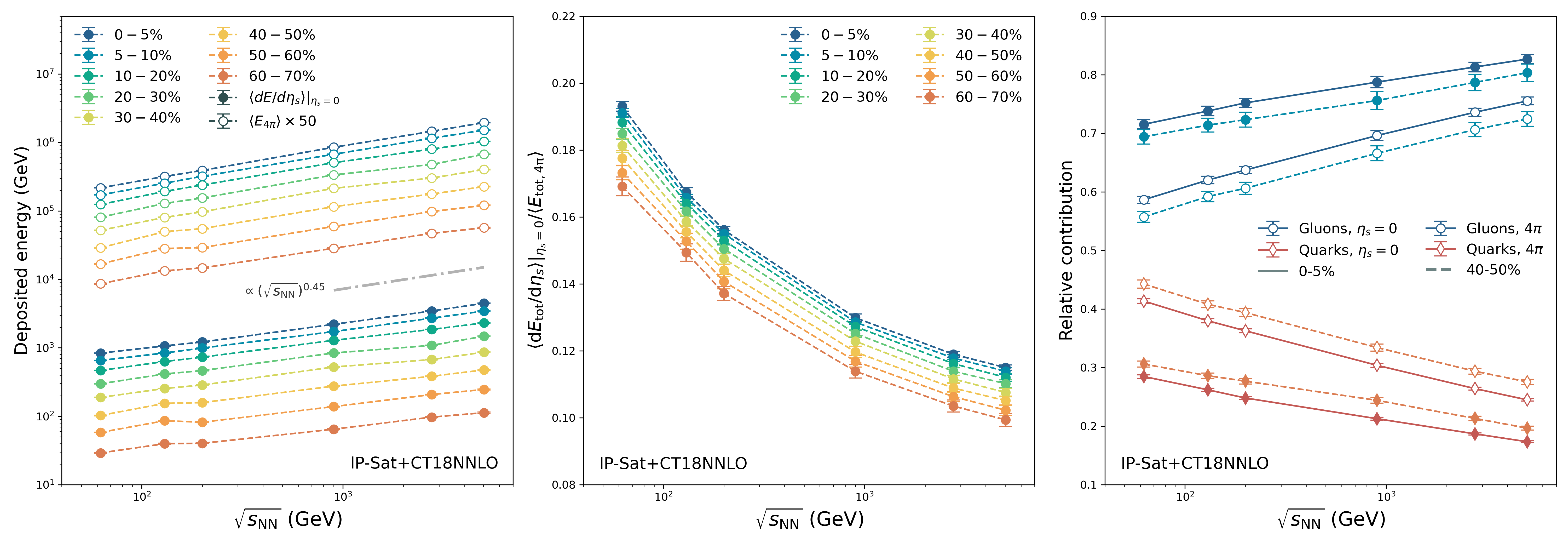}
	\caption{ (\textit{Left}) Energy  deposition for several centralities at midrapidity ($\eta_s =0$) and in $4\pi$ as a function collision energy $\sqrt{s_{\rm NN}}$. (\textit{Center}) Ratio of the total energy deposited at midrapidity normalized with the total deposited energy. (\textit{Right}) Dependence of the fraction of the gluon and quark energy contributions as a function of $\sqrt{s_{\rm NN}}$. Here, the fraction is obtained by normalization to the respective energy at midrapidity (full symbols), as well as the total energy (open symbols). }
	\label{fig:energy}
\end{figure}

In Fig.~\ref{fig:energy} the reader can find several properties of the deposited energy in the \Dipper{}  framework. In the left panel, the excitation function (the $ \sqrt{s_{\rm NN}}$ dependence) of both the midrapidity and total energy are plotted for different centrality classes.  Additionally, the reader can observe that the midrapidity slice and the total energy deposited scale roughly with  $\sqrt{s_{\rm NN}}$ as a power law sufficiently slower than linear increase. The center panel shows the deposited energy at midrapidity as normalized by the full $4\pi$ integrated energy. The ratio exhibits a very minor residual dependence on $\sqrt{s_{\rm NN}}$, from which we can conclude that even though midrapidity energy increases with $\sqrt{s_{\rm NN}}$, more energy is flowing towards the fragmentation regions at higher collisional energies.

In the right panel one can find the relative contribution of quarks (diamonds) and gluons (circles) to the energy deposition. In this panel, contributions taken with respect only to the midrapidity are marked as full symbols, while contributions taken for the whole event are shown as empty markers. For simplicity, only two centralities are shown, $0-5\%$ (line) and  $40-50\%$ (dashed line). In this figure, one can observe that, as intuitively expected from the model, the gluon contribution becomes more important at higher energies. This dependence is more intense for the case of the $4\pi$ integrated energy, which can be understood as the addition of the energy deposited in the fragmentation regions, where quarks with larger $x$ can be easily deflected and deposited into a medium at lower collisional energies. Additionally, the relative contributions exhibit a quite soft dependence on centrality, where the quark contribution becomes more significant for more peripheral events. In the advent of full 3+1D hybrid models, these last findings become quite relevant, especially in the early stages after the initial deposition. Since quark deposition comprises a minimum of 20\% on the current energies available, one can question how much will this inflict the evolution of the thermalizing plasma. These effects may be particularly important in observables such as photon and dilepton production, which have been shown to be sensitive to the process of chemical equilibration of the QGP \cite{Gale:2021emg}, and such effects may be addressed in the future in a quantitative manner using kinetic theory \cite{Garcia-Montero:2023lrd,Du:2020dvp}.

\begin{figure}
	\centering
	\includegraphics[width=0.88\linewidth]{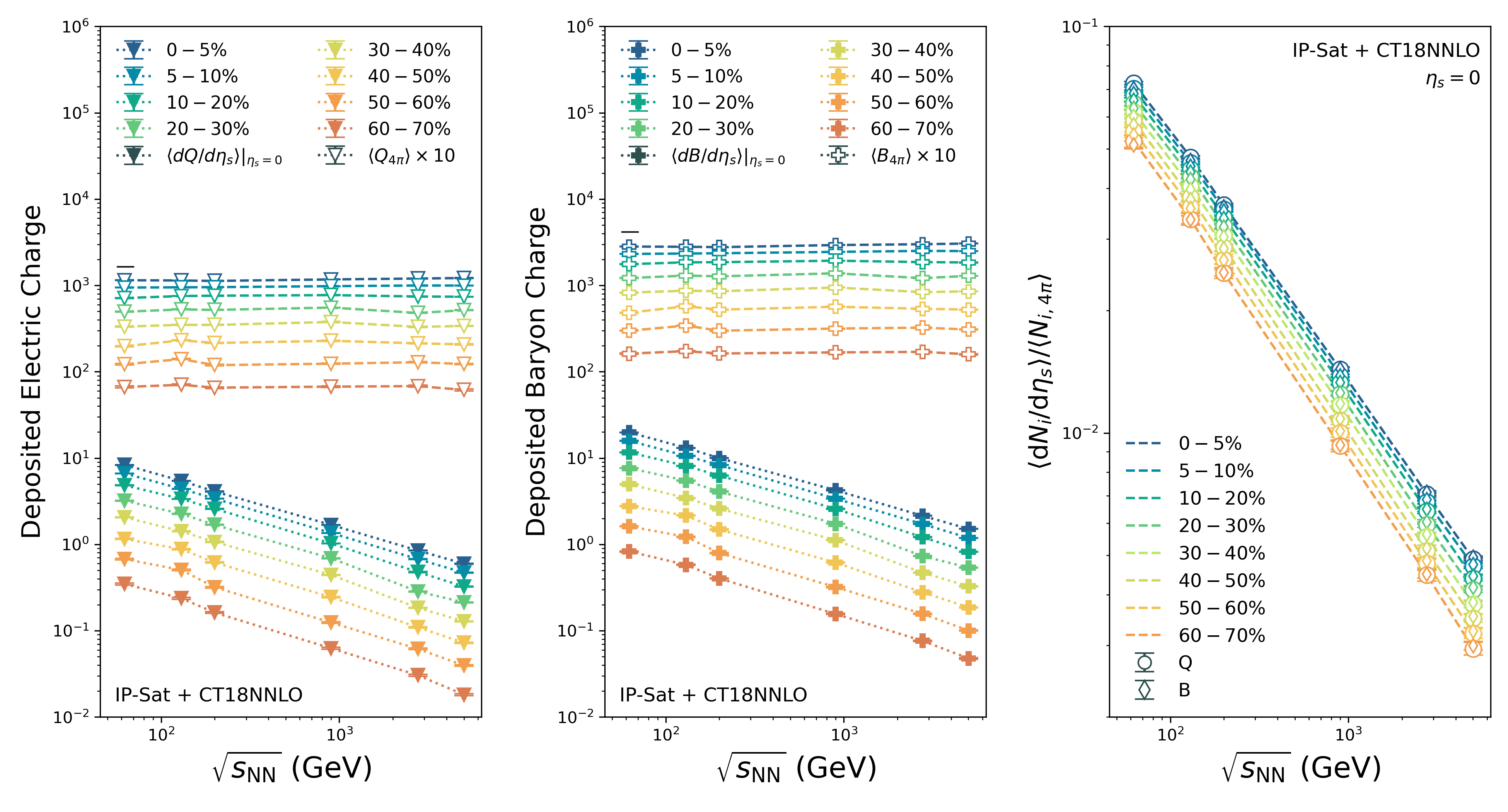}
	\caption{Deposited electric (\textit{left}) and baryon (\textit{center}) charges for different centrality classes as a function of $\sqrt{s_{\rm NN}}$. Shown in full markers are the midrapidity slices, while the empty markers correspond to the total charge deposited with $4\pi$ acceptance. The slim black line is the total incoming electric and baryon charge, $Q_0=2\,Z$ and $B_0=2\,A$.  (\textit{Right}) Ratio of electric and baryon charges at midrapidity to their $4\pi$ integrated corresponding charge. }
	\label{fig:charges}
\end{figure}

In Fig.~\ref{fig:charges} the reader can find several properties of the deposited electric and baryon charges, here denoted as $Q$ and $B$, respectively. In the left and center panels of Fig.~\ref{fig:charges} one can find $Q$ and $B$, respectively, for the midrapidity slice as well as for the total integrated case. The bahavior of these quantitites is qualitatively similar. It is interesting to see that the total integrated charges are virtually constant with respect to $\sqrt{s_{\rm NN}}$ , and their deposition depends uniquely on the geometrical constraints, coming from the sampling of the nucleons from a Woods-Saxon's profile, and results from a strict  constraining of the conserved charges throught the parton PDFs.
On the other hand the charge deposited at midrapidity is monotonically decreasing with $\sqrt{s_{\rm NN}}$. As expected, nuclei become more transparent at midrapidity. The deposition  of charge is shifted with collisional energy more and more towards the fragmentation regions. In the right panel of Fig. \ref{fig:charges} the ratio of the midrapidity and total deposited charge is plotted for both Q and B. The behavior of both charges is identical when comparing the ratios, indicating that the behavior is purely geometrical, as it exhibits exactly the same behavior for electric and baryon charges, regardless of system (Au-Au/ Pb-Pb) or collisional energy.

\section{Conclusions and Outlook}
In this short note we have shown several aspects of the initial energy and charge deposition of a 3D resolved heavy-ion collision in the \Dipper{} framework.  For this we have used the language of the so-called excitation functions to show that the energy deposition presents a soft power-law dependence  on $\sqrt{s_{\rm NN}}$, very similar for both midrapidity and $4\pi$ deposition. Additionally, we have shown the behavior of the  quark and gluon relative contribution with respect to $\sqrt{s_{\rm NN}}$, which exhibits a sharper dependence for the $4\pi$ energy as for the midrapidity slice. 

Additionally, we have shown that the full $4\pi$ deposition of the electric and baryon charges is only geometrical, as it has virtually no dependence on $\sqrt{s_{\rm NN}}$. On the midrapidity slice both charges exhibit a steep decrease with rising energy, which is expected from nuclei becoming more transparent, and the baryon stopping is shifted towards higher values of $\eta_s$. We find that once normalized by their respective full $4\pi$ values of each charge, these ratios exhibit the same values, allowing for a purely geometrical interpretation. For more detailed on the general aspects of the framework and the relevant observables, the reader can find more results on the \Dipper{} framework in the main paper, see ref. \cite{Garcia-Montero:2023gex}. 

Future endeavors for the framework will contain localized fluctuation of charge deposition, which will give not only sub-nucleonic hotspots of electric and baryon charge, but also a locally non-vanishing strangeness distribution. That said, such fluctations should should not affect the results of this work, as the variations should vanish for the expectation values of the charge observables, remaining only at the level of higher order correlations.

\end{document}